\documentclass[a4paper]{jpconf}

\usepackage{graphicx}
\usepackage{citesort}
\newcommand{\ph}{\phantom{1}}

\begin{document}

\title{Gradient correction scheme for bulk and defect positron states in materials: New developments}

\author{J Kuriplach$^1$ and B Barbiellini$^2$}

\address{$^1$ Department of Low Temperature Physics, Faculty of Mathematics and Physics, Charles University in Prague, V Hole\v{s}ovi\v{c}k\'ach 2, CZ-180\,00 Prague, Czech Republic}
\address{$^2$ Department of Physics, Northeastern University, Boston, Massachusetts 02115, USA}

\ead{jan.kuriplach@mff.cuni.cz}

\begin{abstract}
As local density approximation positron calculations systematically underestimate positron lifetimes when they are compared with their experimental counterparts, the generalized gradient approximation (GGA) for positrons was introduced in the 1990s,  in analogy with the GGA for electrons.
New developments in the GGA for positrons are summarized and presented here and it is also discussed how they affect and possibly improve calculated positron lifetimes.
In particular, these new GGA approaches are based on the recent perturbed hypernetted-chain and quantum Monte Carlo results.
\end{abstract}

\section{Introduction}

The bulk positron lifetime and the affinity are fundamental positron characteristics in materials. 
They are accessible experimentally and they can be calculated on 
the basis of the electronic structure of the examined materials. 
An important aspect of such computations is the type of approach to the electron-positron interactions for positrons experiencing the environment of densely arranged atoms in materials. Though the interaction of a positron with nuclei is of purely electrostatic origin, the interaction of the same positron with electrons requires the consideration of electron-positron correlations in addition to electrostatics. 
First attempts to describe such complex electron-positron interaction were based on a local density approximation (LDA) using a theory for a positron embedded in a homogeneous electron gas \cite{AP}. 
However, it turned out that, in general, LDA approaches underestimate positron lifetimes, thus some corrections beyond the LDA for positrons are necessary. 
In analogy with electrons where gradient corrections to the LDA are often successful to cure various LDA deficiencies, a gradient correction scheme for electron-positron correlations 
was introduced almost 20 years ago \cite{gga1,gga2}. 
This scheme called generalized gradient approximation (GGA) was quite successful in removing the shortcomings of original positron LDA theories, but still some problems remained (for example, the aluminum calculated lifetime was too long compared to the experiment). The main effect of the GGA was identified to be the reduction of the enhancement factor for core and other more localized electrons.

Here, we present a refinement \cite{gga4} of the original GGA for positrons. 
For this purpose, we use the results of a recent quantum Monte Carlo (QMC) 
study \cite{qmc} for the homogeneous electron gas with one positron impurity, 
where accurate parametrizations of the corresponding LDA electron-positron 
enhancement factor and correlation energy as functions of the electron density were determined. 
The positron gradient correction is then applied to this LDA based on QMC simulations. 
We show the results of positron lifetime calculations for selected materials (both metals and semiconductors) and we compare them to available experimental data. 
We can observe an overall improvement of the agreement between theory and experiment compared to the original GGA scheme. 
We also compare our new GGA approach to that developed recently by Boro\'nski \cite{ggaboro}, which is based on the perturbed hypernetted chain (PHNC) approximation \cite{phnc}, and we also notice the similarity between the results obtained with both approaches, except in the case of the alkali metals. 
For our positron calculations, we make use of precise, self-consistent electronic 
structures computed without any shape approximation for the electron density and potential. 
This is important to obtain numerically accurate positron properties. 
Finally, we demonstrate the capabilities of the new GGA scheme for defect studies 
and present the positron lifetime and binding energy results for a 
monovacancy in several materials.

\section{Theoretical background and computational methods}

The enhancement factor describes the pile-up of electrons around the positron.
Enhancement factors are usually expressed in terms of a power series of the density parameter ($r_s$) where some exponents can be non-integer, like the widely used Boro\'nski and Nieminen (BN) enhancement \cite{tdft}.
In any case, the simplest form to reproduce/fit the results of many-body calculations of the enhancement is the following expression \cite{gga1}
\begin{equation}
\gamma(r_s) = 1 + 1.23\,r_s + p\,r_s^2 + r_s^3/3 \,,
\label{eq:gamma}
\end{equation}
where $p$ is the only adjustable parameter, and the linear and cubic terms are fixed (see \cite{gga1} for details).
The density parameter is a radius of a sphere containing one electron, i.e. $4\pi\,r_s^3\,\rho^-/\,3 = 1$, $\rho^-$ being the electron density.
The fit of QMC data \cite{qmc} gives $p=-0.22$ whereas the original Arponen and Pajanne (AP) data \cite{AP} results in $p=-0.0742$. 
Interestingly, the Stachowiak and Lach's (SL) enhancement within the PHNC approximation \cite{phnc} is best reproduced with $p=-0.1375$.

\begin{figure}[h]
\begin{minipage}{7.7cm}
\includegraphics[width=7.7cm,clip=true]{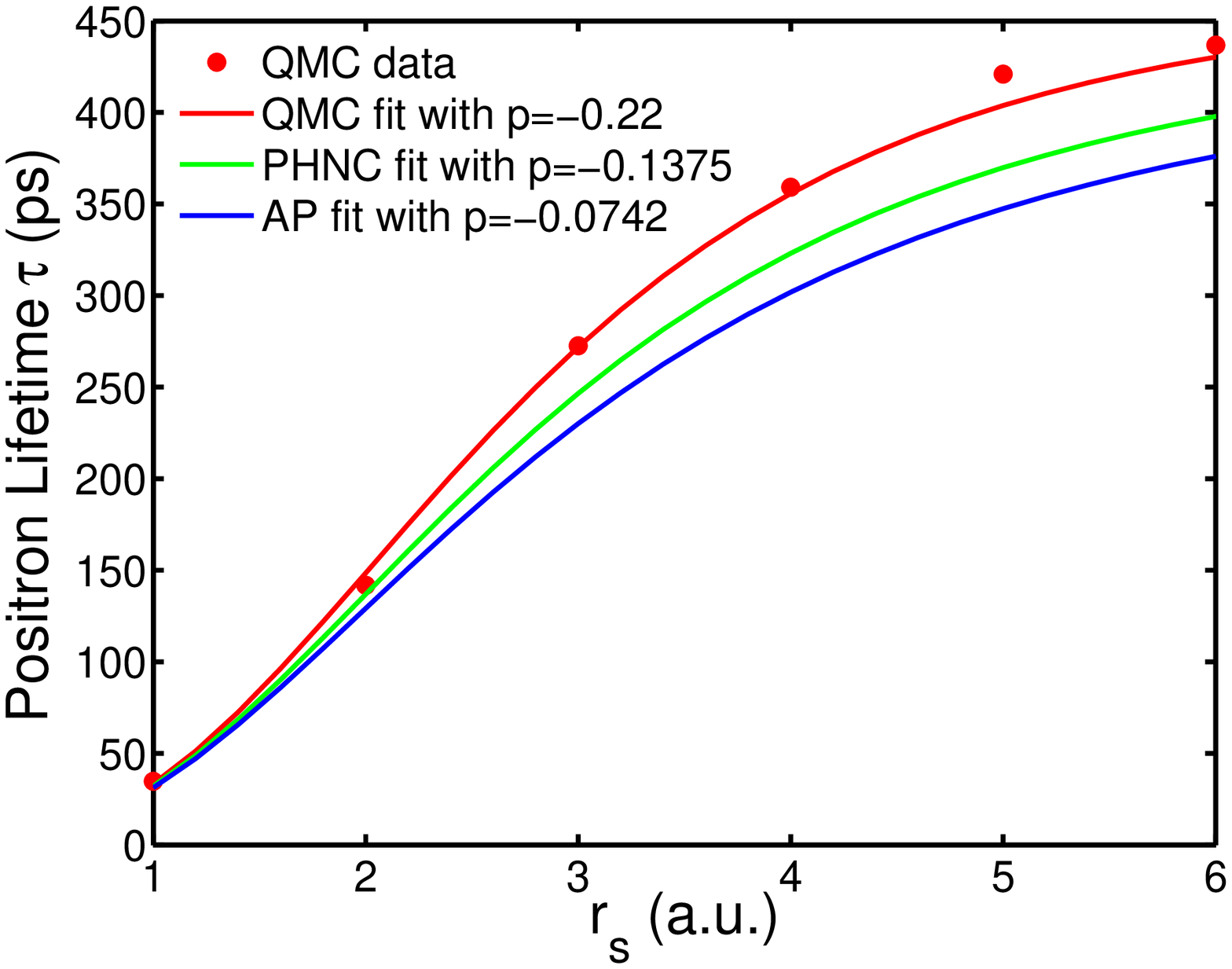}
\caption{Positron lifetime $\tau$ as a function of density parameter $r_s$ for the homogeneous electron gas. Lifetimes calculated using the fitted parameter $p$ in equation (\ref{eq:gamma}) for several approaches are shown.}
\label{f:tau}
\end{minipage}\hspace{5mm}%
\begin{minipage}{7.8cm}
\vspace*{-4.3mm}
\includegraphics[width=7.7cm,clip=true]{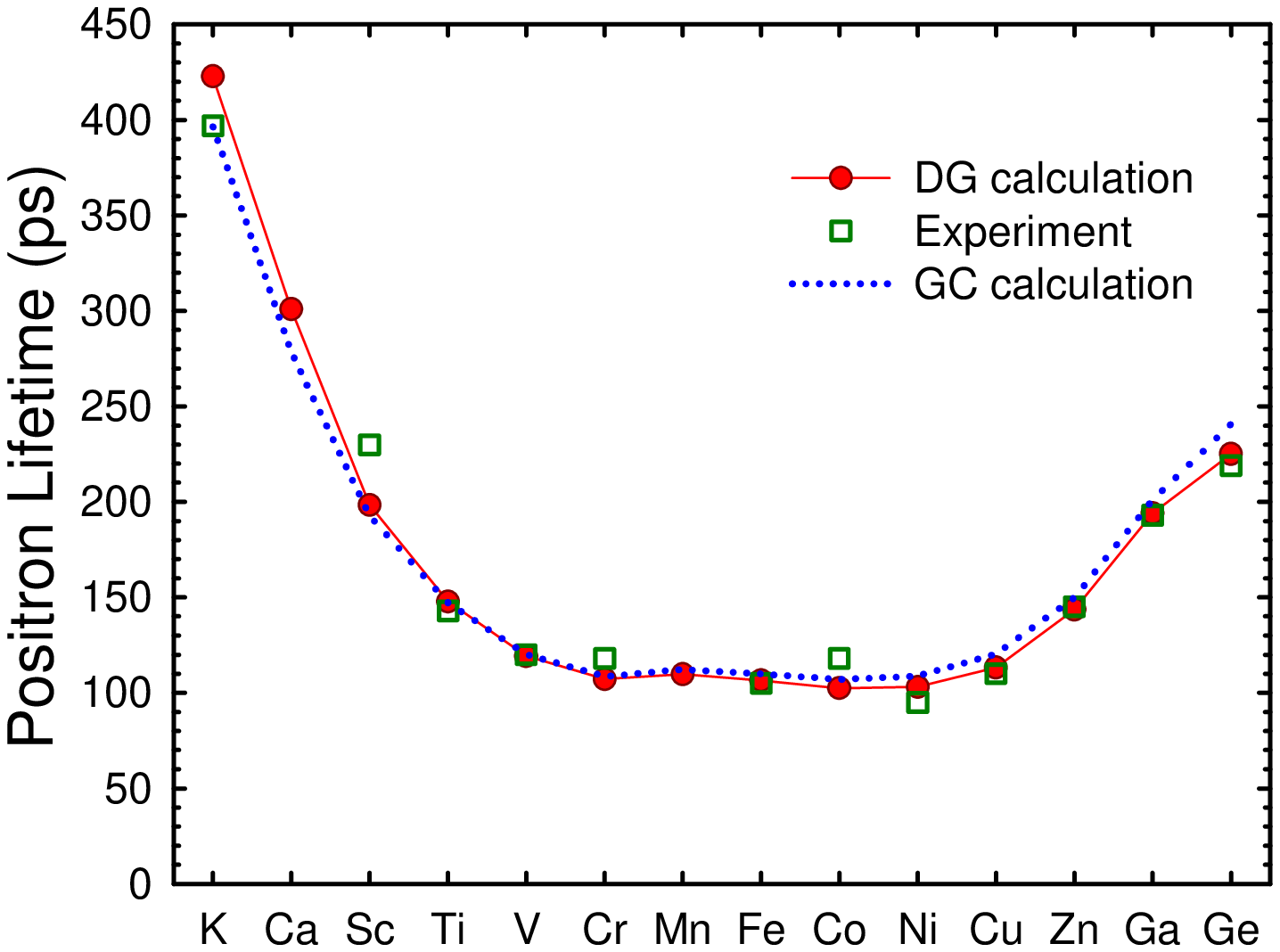}%
\vspace*{3mm}
\caption{Positron lifetimes for the 4th period of the periodic table of elements. Both calculated (DG and GC approaches) and experimental values are plotted.}
\label{f:4p}
\end{minipage}
\end{figure}

Fig. \ref{f:tau} shows the dependence of the positron lifetime ($\tau$) on the density parameter calculated using the just mentioned LDA enhancements as the inverse of the annihilation rate
\begin{equation}
\lambda = 1/\tau = \pi r_e^2 c\,\rho^-(r_s)\,\gamma(r_s) \,.
\label{eq:lambda}
\end{equation}
In this equation, $r_e$ is the classical electron radius and $c$ is the speed of
light. 
One can see that there are quite large differences among approaches investigated, especially for smaller densities (larger $r_s$), which results in different positron lifetimes for real materials, as demonstrated below.
The QMC enhancement \cite{qmc} (denoted in the following by abbreviation DB) exhibits the longest lifetimes compared to other two approaches: PHNC (or SL) and AP.

The GGA for positrons is introduced using the equation
\begin{equation}
\gamma_{GGA} - 1 = (\gamma_{LDA}-1) \exp(-\alpha \epsilon) \,,
\end{equation}
where $\gamma_{GGA}$ and $\gamma_{LDA}$ are, respectively, GGA and LDA versions of the enhancement factor. 
The $\epsilon$ quantity characterizes the density gradient and $\alpha$ is an adjustable parameter.
In practical calculations, $\epsilon$ is chosen to be 
$(\ln \nabla \rho^- / q_{TF})^2$ where $1/q_{TF}$ is the Thomas-Fermi screening length (see \cite{gga1} for details).
Thus, the original GGA for positrons (denoted further by GC) based on the AP approach was used with $\alpha=0.22$ \cite{gga1,gga2}.
Boro\'nski recently suggested that the SL theory should be used with $\alpha=0.10$ \cite{ggaboro} (marked as SG in the following). 
And here we suggest \cite{gga4} that $\alpha=0.05$ should be used with the QMC (DB) approach (abbreviated as DG hereinafter).
In the following, we show how these GGA approaches compare with each other and how they match experimental lifetime results.

The electronic structure calculations for selected materials were carried out using the WIEN2k code \cite{wien2k} which is an implementation of the augmented plane wave plus local orbital (APW+lo) method \cite{APWlo}.
This method is considered to be one of the most accurate methods to calculate electronic structure of solids.
The WIEN2k program also performs full-potential calculations, which impose no shape restrictions for the electron density and potential.
The LDA electron exchange-correlation functional based on QMC simulations
by Ceperley and Alder \cite{eCAXC} and parametrized by Perdew and Wang \cite{eCAPWXC} was utilized in the course of the electronic structure calculations.

The positron wave function and energy were obtained using a computer code developed on the basis of a finite difference method \cite{ATSUP1,ATSUP2} to solve
the positron Schr\"odinger equation.
The positron potential was constructed using the electron Coulomb potential and a positron correlation potential obtained from the total electron density.
Both the Coulomb potential and electron density were taken from self-consistent WIEN2k electronic structure calculations described above.
The positron correlation potential in GGA calculations was obtained by scaling \cite{gga1} the corresponding LDA potential.
The positron potential and wave function are calculated on a regular 3D mesh, which covers an appropriately chosen crystallographic unit cell of the studied material.
The solver uses a numerical procedure based on the conjugate gradient method and it has been previously used within a non-self-consistent atomic superposition scheme \cite{ATSUP2}.
The positron lifetime is calculated using equation (\ref{eq:lambda}) by multiplying its right hand side by the positron density and integrating over the unit cell used 
in calculations.
Various numerical parameters of the WIEN2k code as well as of the positron code were checked and optimized in order to obtain accurate electronic structures and positron wave functions to perform precise positron calculations (the details are described in \cite{gga4}).
As a result, the lifetimes presented in this work are calculated within the precision 0.1 ps.
In this way, the comparison with experimental lifetime data is well warranted.
It is also possible to test an influence of underlying physical approximations (like electron exchange-correlations) on the positron characteristics with negligible disruption of numerics.

\section{Results and discussions}

In order to demonstrate the capabilities of the new QMC enhancement and its gradient corrected version, we calculated positron lifetimes for selected elements and compounds.
Namely, these are alkali metals, elements of the 4th period of the periodic table and several oxides which attract attention at present.

\begin{table}[b]
\caption{Lifetimes (in ps) for alkali metals calculated using various approaches to electron-positron correlations. The third column lists the average $r_s$ parameter (in a.u.) based on one conduction $s$-electron per atom.
The last column (Exp.) presents experimental lifetimes from \cite{expalkali}.
The results for Li and Na were taken from \cite{gga4}.}
\begin{center}
\begin{tabular}{llccccccccc}
\br
Metal & Structure & $r_s$ & BN & AP & GC & SL & SG & DB & DG & Exp. \\
\mr
Li  & bcc& 3.27 & 299.5 & 259.4 & 283.4 & 277.2 & 295.0 & 303.5 & 315.8   & 291 \\
Na  & bcc& 3.99 & 328.1 & 290.7 & 338.6 & 309.9 & 341.6 & 342.8 & 364.2   & 338 \\
K   & bcc& 4.96 & 373.7 & 338.5 & 396.4 & 360.4 & 395.0 & 400.6 & 422.8   & 397 \\
Rb  & bcc& 5.20 & 376.6 & 342.0 & 406.3 & 364.1 & 400.5 & 403.7 & 426.4   & 406 \\
\br
\end{tabular}
\label{t:alk}
\end{center}
\end{table}

Alkali metals represent interesting test systems in positron calculations because their conduction electrons behave almost as free, the corresponding electron densities being very low compared e.g. to transition metals.
The results of lifetime calculations are presented in table \ref{t:alk}.
In the third column, we also show the average $r_s$ parameter. 
The positron lifetime increases with increasing $r_s$ (see equation (\ref{eq:lambda})), i.e. from Li to Rb.
The effect of the gradient correction is quite large, especially for the original GGA, and can be attributed to core electrons (see \cite{gga4}).
The best agreement with experiment \cite{expalkali} can be found for GC, LG and DB approaches, the DG values being too long.
However, it should be mentioned that these experiments were carried out about 45 years ago and the results might not be very precise considering that no source correction was done, and other effects like the presence of vacancies \cite{seeger} and impurities and thermalization issues \cite{thermalization} could influence measured lifetimes.
Therefore, it would be very desirable to perform new lifetime measurements for alkalies addressing these possible problems.

Being inspired by recent calculations of positron lifetimes for elements of the periodic table \cite{campillo}, we have determined positron lifetimes for the elements of the 4th period.
The results are presented in table \ref{t:4g}. 
Fe, Co and Ni were treated as ferromagnetic in calculations whereas an antiferromagnetism of Cr \cite{Crmag} and Mn \cite{Mnmag} was not considered.
Experimental lifetimes are also given in table \ref{t:4g} and are taken mostly from reviews \cite{seeger,perexp03,sourceEldrup}.
To the best of our knowledge, Ca and Mn have not been investigated by means of positron lifetime spectroscopy so far.
In table \ref{t:4g}, we give most of measured lifetime values in the form ``X+'',
where X is the minimum bulk lifetime detected in experiment.
In fact, there is quite big scatter in experimental data for many elements -- for instance, lifetimes for Cu typically vary in the (110 -- 120) ps range -- and specifying the lower limit gives some idea about this uncertainty.
The most probable origin of such uncertainties is the source correction which varies among positron laboratories performing measurements (see e.g. \cite{sourceDundee}).

\begin{table}[h]
\caption{Lifetime results (in ps) for the 4th period elements (atomic numbers 19--32) calculated using various approaches to electron-positron correlations.
The last column indicates experimental values.
The results for Fe, Cu and Zn were taken from \cite{gga4}.}
\begin{center}
\begin{tabular}{llccccccccc}
\br
Element & Structure & BN & AP & GC & SL & SG & DB & DG & Exp. \\
\mr
K   & bcc       & 373.7 & 338.5 & 396.4 & 360.4 & 395.0 & 400.6 & 422.8  & 397 & \\
Ca  & fcc       & 288.5 & 250.0 & 278.3 & 267.0 & 283.8 & 290.7 & 301.0  & --  & \\[1mm]
Sc  & hcp       & 195.1 & 171.3 & 192.6 & 181.7 & 193.7 & 191.6 & 198.3  & 230 & \\
Ti  & hcp       & 144.4 & 129.7 & 147.0 & 136.8 & 146.2 & 142.7 & 147.8  & 143+ & \\
V   & bcc       & 115.1 & 105.4 & 120.1 & 110.5 & 118.4 & 115.1 & 119.3  & 120+ & \\
Cr  & bcc       & 102.6 &\ph94.8& 108.5 &\ph99.2& 106.5 & 103.4 & 107.2  & 118+ & \\
Mn  & $\alpha$-Mn, A12 &
                  104.7 &\ph96.7& 112.1 & 101.2 & 109.3 & 105.4 & 109.7  & --  &  \\
Fe  & bcc       & 101.1 &\ph93.6& 109.9 &\ph97.9& 106.5 & 102.1 & 106.5  & 105+ & \\
Co  & hcp       &\ph96.6&\ph89.8& 106.9 &\ph93.8& 102.7 &\ph97.8& 102.5  & 118+ & \\
Ni  & fcc       &\ph96.9&\ph90.2& 108.8 &\ph94.1& 103.9 &\ph98.2& 103.2  &\ph95+ &\\
Cu  & fcc       & 106.4 &\ph98.4& 120.3 & 102.9 & 114.4 & 107.3 & 113.2  & 110+ & \\
Zn  & hcp       & 137.1 & 124.3 & 149.8 & 130.7 & 144.2 & 136.4 & 143.6  & 145+ & \\[1mm]
Ga  & $\alpha$-Ga, A11 &
                  187.0 & 166.2 & 200.9 & 175.9 & 193.2 & 184.8 & 194.0  & 193+ & \\
Ge  & diamond   & 214.2 & 189.8 & 240.8 & 201.2 & 225.2 & 212.8 & 225.2  & 219+ & \\
\br
\end{tabular}
\label{t:4g}
\end{center}
\end{table}

The lifetime trend across the 4th period can be clearly observed in figure \ref{f:4p}.
There is a clear minimum in the calculated lifetimes around the center of the period, which corresponds to heavier 3$d$ metals having the largest electron densities (smallest volumes per atom \cite{campillo}).
Such a trend was first observed by Brandt {\it et al.} \cite{Brandt75} (see also \cite{campillo}) and here it is confirmed by more precise calculations. 
This trend is also seen for experimental values though it is not that apparent due to the scatter in experimental lifetimes.
Table \ref{t:4g} also compares the original GGA (GC) with its new version based on QMC results (DG).
One can see that the agreement with experiment is better for the DG approach. 
Moreover, the GC approach gives systematically lower lifetimes till Ti whereas in the remaining part of the period the lifetimes are longer.
The observed trend through the period clearly shows that positron lifetime spectroscopy is well capable of distinguishing between $sp$ and transition elements.

At the beginning of the period, there are two $s$ metals (K and Ca) having large atomic volumes (and small electron densities), which results in very long lifetimes ($>$300 ps).
With the increasing atomic number the $d$-block with lighter metals starts, and the atomic volume decreases (electron density increases) in the series, which is reflected in a decrease of the positron lifetime.
The last element in this part of the 4th period is Mn having quite complicated cubic structure, which slightly disturbs the decreasing trend.
At the beginning of heavier 3$d$ metal series, the lifetimes are the shortest with an increase towards the end due to hcp Zn.
The next two elements considered, Ga \cite{Gaexp} and Ge, confirm the increasing lifetime trend because of their quite open structures (A11 and diamond).
Considering the most frequently studied elements Fe and Cu, the lifetimes calculated using the DG (and also SG) approach are very close to experimental ones apart from the scatter mentioned.

Finally, we present the calculated positron lifetimes for several oxides which are frequently studied at present (table \ref{t:oxides}).
So far, we have discussed elements and for these cases metallic or covalent bonding takes place.
On the other hand, oxides exhibit ionic bonding (in addition to covalent one) and a large charge transfer occurs between oxygen and other atoms forming the oxides.
These effects also influence positron properties since positrons are attracted to oxygen anions and are repelled from cations though it cannot be viewed as a bound state of positrons with oxygen anions; positrons certainly reside in the interstitial space among anions and cations.

\begin{table}[thb]
\caption{Positron lifetime results (in ps) for selected oxides. Various approaches to electron-positron correlations have been used.
The results for MgO and and ZnO were taken from \cite{gga4}.}
\begin{center}
\begin{tabular}{llccccccccc}
\br
Oxide & Structure & BN & AP & GC & SL & SG & DB & DG & Exp. \\
\mr
MgO & rock salt & 117.8 & 108.7 & 145.4 & 113.8 & 131.8 & 119.0 & 128.4  & 130 & \\
SiO$_2$ & $\alpha$-quartz &
                  205.8 & 186.4 & 297.1 & 196.7 & 246.9 & 207.7 & 233.2  & 
$\sim$261 & \\ 
Cu$_2$O &cuprite& 147.2 & 133.8 & 177.2 & 140.6 & 161.9 & 147.2 & 158.3  & $\sim$174&  \\
ZnO & wurtzite  & 143.9 & 131.6 & 183.3 & 138.2 & 162.3 & 144.6 & 156.9  & 151+& \\
ZrO$_2$&fluorite& 128.2 & 118.0 & 158.8 & 123.7 & 142.5 & 128.9 & 138.4  & $\sim$140& \\
CeO$_2$&fluorite& 138.1 & 126.7 & 172.6 & 133.0 & 154.0 & 138.5 & 149.2  & ${<}187$& \\
\br
\end{tabular}
\label{t:oxides}
\end{center}
\end{table}

In general, experimental positron studies of oxides are rather prone to effects of impurities.
For instance, MgO was shown to be affected by unintentional Ga doping \cite{MgOdoping} which results in the creation of Mg vacancies.
When the doping level is decreased under certain limit, the effect disappears and the measured bulk lifetime (see e.g. \cite{MgOPAS} and references therein) agrees well with calculations, as documented in table \ref{t:oxides}.
Another example is ZnO where H (\cite{Brauer09}) and Li (\cite{Johansen11}) doping plays an important role.
Current results (SG and DG) suggest that the lifetime for bulk ZnO is around 160 ps.
Furthermore, silica in its $\alpha$-quartz (or low quartz) form was also studied.
Its experimental bulk lifetime is not very well known and was estimated to be about 260 ps in \cite{alphaSiO2}.
The calculated lifetimes are apparently shorter (except the GC approach) and this indicates that the mentioned experimental value might require a revision and thereby a new experiment.
Cuprite (Cu$_2$O) is one of the most studied oxides with various practical applications.
Like for $\alpha$-quartz, the bulk lifetime of Cu$_2$O is not well fixed and recent experiment \cite{Cu2Oexp} gives an estimate of 174 ps, which in the light of present results (see table \ref{t:oxides}) appears to be a bit too long lifetime.
Zirconia (ZrO$_2$), one of the most studied ceramic materials, needs stabilization -- e.g. by yttria -- to keep the fluorite structure at ambient conditions.
The bulk lifetime of ZrO$_2$ was measured to be approximately 141 ps in \cite{ZrO2exp} and the SG and DG results are well in line with this experimental value.
Finally, ceria (CeO$_2$) exhibiting the same structure as zirconia has many practical applications.
It has been only rarely investigated by means of positron annihilation.
Therefore, its bulk positron lifetime is not fixed and an estimate based on a study  of CeO$_2$ nanoparticles presented in \cite{CeO2exp} gives an upper limit of 187 ps, as the shortest defect component detected in samples. 
Though the authors claim that this component corresponds to a reduced bulk lifetime, it is not very likely when the lifetime results from table \ref{t:oxides} are consulted.

Concerning defects studies, we refer readers to our work \cite{gga4} where we show first results for vacancies in Al, Si and Cu obtained with new GGA enhancements.

\section{Conclusions}

We have calculated positron lifetimes in selected elements and compounds based on reliable first principles electronic structure calculations taking into account highly precise electronic charge transfers due to an electron self-consistent full potential.
The LDA for positrons does not take into consideration charge inhomogeneities due to non metallic charge distribution and the effect of the electron-nuclear interaction which is disrupting the pile-up of electronic charge around the position of the positron.
Therefore, we show that gradient corrections to the LDA are still needed in such circumstances despite their intensity is reduced in comparison to the original GGA scheme \cite{gga1}.
On the basis of presented calculations, it is currently difficult to decide -- by making comparisons with available experimental positron lifetimes -- which is the best GGA approach among the GGA-PHNC (SG) and the GGA-QMC (DG). Thus, further and more precise experiments are highly desirable to answer this important question.
Furthermore, we plan to test the new positron GGA's in calculations of the momentum distribution of electron-positron pairs and see whether we can obtain some improvements in addressing the problems outlined e.g. in \cite{Laverock}.

\section*{Acknowledgments}
The authors would like to thank M.~Eldrup, D.~Keeble and I.~Proch\'azka for helpful discussions concerning various aspects of positron lifetime measurements.
N.D.~Drummond is acknowledged for providing us with numerical results of QMC calculations.
This research used resources of the National Supercomputing Center IT4Innovations (Czech Republic), which is supported by the Op VaVpI project number CZ.1.05/1.1.00/02.0070.
Partial support by the Czech Science Foundation (project P108/12/G043) is appreciated.
B.B. is supported by the US Department of Energy, Office of Science, Basic Energy Sciences Contract No. DE-FG02-07ER46352. He has also benefited from Northeastern University's Advanced Scientific Computation Center (ASCC), theory support at the Advanced Light Source, Berkeley, and the allocation of computer time at NERSC through Grant No. DE-AC02-05CH11231.

\section*{References}
\bibliography{ggam}

\end{document}